\documentclass[twocolumn,twoside,slac_two]{revtex4}
\usepackage{graphicx}
\graphicspath{{./}{./figures/}}
\usepackage{fancyhdr}
\newcommand{\ep}{\epsilon}
\newcommand{\MSbar}{\overline{\mbox{MS}}}
\newcommand{\msbar}{\overline{\mbox{\scriptsize MS}}}
\newcommand{\ARI}{\mbox{RI/MOM}}
\newcommand{\ari}{\mbox{\scriptsize RI/MOM}}
\newcommand{\RIp}{\mbox{RI}^\prime\mbox{/MOM}}
\newcommand{\rip}{\mbox{\scriptsize RI}^\prime\mbox{\scriptsize/MOM}}
\newcommand{\RISMOM}{\mbox{RI/SMOM}}
\newcommand{\RISMOMB}{\mbox{RI/SMOM}_{\gamma_{\mu}}}
\newcommand{\rismomb}{{\scriptsize{\mbox{RI/SMOM}_{\gamma_{\mu}}}}}
\newcommand{\rismom}{\mbox{\scriptsize RI/SMOM}}
\newcommand{\als}{\alpha_s}
\newcommand{\Tr}{\mbox{Tr}}
\newcommand{\Fslash}[1]{\!\not{\hbox{\kern-2pt ${#1}$}}}
\newcommand{\unitop}{1{\hbox{\kern-6.4pt $1$}}}
\newcommand{\CF}{C_F}
\newcommand{\CA}{C_A}
\newcommand{\TF}{T_F}
\newcommand{\nf}{n_f}

\begin{document}

\title{Improving the precision of light quark mass determinations}
\author{Christian Sturm}
\affiliation{Physics Department, Brookhaven National Laboratory, Upton, NY 11973, USA}
\begin{abstract}
We discuss the concepts and the framework of the renormalization
procedure in regularization-invariant momentum subtraction
schemes. These schemes are used in the context of lattice simulations
for the determination of physical quantities like light quark masses. We
focus on the renormalization procedure with a symmetric subtraction
point of a quark mass and discuss its conversion from the $\RISMOM$ scheme
to the $\MSbar$ scheme. A symmetric subtraction point allows for a
lattice calculation with a reduced contamination from infrared effects. The
perturbative series for the resulting matching factor at next-to-leading
order is better behaved for the symmetric momentum configuration
than for the exceptional one which can decrease the uncertainty in
light quark mass determinations.
\end{abstract}
%%
%%%%%%%%%%%%%%%%%%%%%%%%%%%%%%%%%%
%%
\maketitle
\thispagestyle{fancy}
%%
%%%%%%%%%%%%%%%%%%%%%%%%%%%%%%%%%%
%%
\section{Introduction}
In continuum perturbation theory dimensional
regularization~\cite{tHooft:1972fi} is for many phenomenological
applications the most common regulator for divergent
loop-integrals. Physical observables computed to higher orders are often
conveniently renormalized in the minimal subtraction ($\MSbar$)
scheme~\cite{tHooft:1973mm,Bardeen:1978yd}; thus the renormalization
constants are well known to high order in perturbation theory.
%%%

However, the $\MSbar$ scheme is fixed to dimensional regularization and
is usually not directly applicable in lattice calculations. In the
context of lattice simulations in combination with non-perturbative
renormalization, regularization-invariant momentum subtraction(MOM)
schemes are used~\cite{Martinelli:1994ty}. In
regularization-invariant(RI) schemes the renormalization constants are
often fixed by demanding that the renormalized quantity is equal to its
lowest order value for a special momentum configuration(subtraction
point). In order to convert results computed in a $\ARI$ scheme to the
$\MSbar$ scheme matching factors are needed which translate an
observable renormalized in the given $\ARI$ scheme into the $\MSbar$
scheme. These matching factors can be computed perturbatively. For
example, the computation of the mass conversion factor $C^{\ari}_m$,
which turns a quark mass renormalized in the $\ARI$ scheme into the
$\MSbar$ scheme, or the conversion factor $C^{\ari}_q$, which performs
the corresponding conversion of the quark fields, are both known up to
three-loop order in perturbative Quantum Chromodynamics(QCD)~\cite{
Martinelli:1994ty, Franco:1998bm, Chetyrkin:1999pq}. In the $\RIp$
scheme which is also used in lattice simulations these matching factors
are also known to the same order~\cite{Chetyrkin:1999pq,Gracey:2003yr}.
%%%

Apart from quark masses, also the strong coupling constant $\alpha_s$
has been considered in momentum subtraction
schemes~\cite{Celmaster:1979km,Braaten:1981dv,Jegerlehner:1998zg,Chetyrkin:2000fd,Chetyrkin:2000dq,Chetyrkin:2008jk}.
Taking into account vertex diagrams, one has a larger choice of defining
the subtraction point at which the renormalization constants are fixed
through different momentum configurations. Two examples are the
exceptional(or asymmetric) subtraction point with the external
minkowskian momenta $p_1^2=p_2^2=-\mu^2$, $p_3=0$ or the
non-exceptional(or symmetric) momentum configuration with
$p_1^2=p_2^2=p_3^2=-\mu^2$. The symbol $\mu$ denotes here the
renormalization scale.
%%%

In Ref.~\cite{Sturm:2009kb} the renormalization of quark bilinear
operators for a symmetric momentum subtraction point has been studied in
the $\RISMOM$ scheme for the vector, axial-vector, scalar, pseudo-scalar
and tensor operators, where the S in SMOM stands for symmetric. The
corresponding renormalization constants $Z$ are related among each other
through the Ward-Takahashi identities. These relations have been used to
perform mass renormalization with a symmetric subtraction point.
%%%

Light up-, down- and strange-quark masses were determined in
Ref.~\cite{Allton:2008pn} through lattice simulations in the $\ARI$
scheme and subsequently converted to the $\MSbar$ scheme. However, the
perturbative expansion of the matching factor $C^{\ari}_m$ in QCD shows
poor convergence and leads currently to a significant contribution to
the error on these quark masses. The uncertainty from the
renormalization procedure of the quark masses in
Ref.~\cite{Allton:2008pn} amounts to around 60\% of the total error.  In
order to renormalize the quark masses in the lattice simulation, the
renormalization constants need to be determined in a lattice
calculation~\cite{Aoki:2007xm}. As shown in Ref.~\cite{Aoki:2007xm}, a
symmetric subtraction point involves non-exceptional momenta which imply
a lattice simulation with suppressed contamination from infrared effects,
whereas the asymmetric subtraction point is less suited, since effects
of chiral symmetry breaking vanish only slowly for large external
momenta as a result of Weinberg's theorem~\cite{Weinberg:1959nj}.
%%%

This proceedings contribution is organized as follows: in
Section~\ref{sec:GenAndNot} we introduced the used conventions and
notations and discuss generalities. In Section~\ref{sec:results} we
summarize the results for the mass renormalization with a symmetric
subtraction point, addressing the matching factors and the two-loop
anomalous dimensions. Finally we close with the conclusions in
Section~\ref{sec:conclusion}.
\section{Generalities and Notation\label{sec:GenAndNot}}
\begin{figure}[!ht]
\begin{center}
\begin{minipage}{2.5cm}
\includegraphics[bb=141 617 248 644,width=2.5cm]{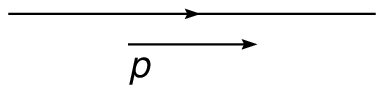}
\vspace{-1cm}
\begin{center}
{\scriptsize{(a)}}
\end{center}
\end{minipage}
\hspace{-0.1cm}
\begin{minipage}[t]{4cm}
\includegraphics[bb=71 618 248 722,width=4cm]{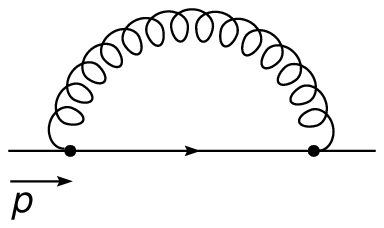}
\vspace{-1cm}
\begin{center}
\hspace{1.35cm}{\scriptsize{(b)}}
\end{center}
\end{minipage}
\end{center}
\vspace{-0.3cm}
\caption{\label{fig:propagator}
QCD corrections up to one-loop order to the quark propagator.}
\end{figure}

Bare and renormalized quantities are related through the renormalization
constants $Z$, like the renormalization constant $Z_m$ of the quark mass
$m$ and the renormalization constant $Z_q$ of the fermion field~$\Psi$
with the properties
\begin{equation}
m_R=Z_m\*m_B\quad\mbox{and}\quad
\Psi_R=Z_{q}^{1/2}\*\Psi_B.
\end{equation}
The subscripts {\scriptsize{$R$}} and {\scriptsize{$B$}} denote
renormalized and bare objects. These renormalization constants can be
fixed through renormalization conditions imposed on self-energy type
diagrams as shown in Fig.~\ref{fig:propagator} given by
\begin{eqnarray}
-i\*S(p)&=&\int\!dx e^{ipx}\langle T[\Psi(x)\overline{\Psi}(0)]\rangle\\
        &=&{i\over \Fslash{p}-m+i\*\ep-\Sigma(p)},
\end{eqnarray}
where $\Sigma(p)$ incorporates the higher order QCD corrections.\\
\\
In perturbation theory the conditions for the self-energies in the last
line of Table~\ref{tab:Cond} fix the renormalization constants in the
$\ARI$ scheme.
%%%

\begin{figure}[!ht]
\begin{center}
\begin{minipage}{2.5cm}
\includegraphics[bb=118 588 226 679,width=2.5cm]{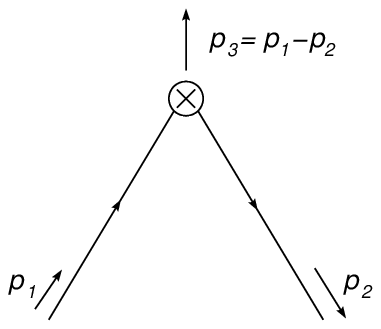}
\vspace{-1cm}
\begin{center}
{\scriptsize{(a)}}
\end{center}
\end{minipage}\\[0.5cm]
\begin{minipage}{2.5cm}
\includegraphics[bb=118 588 226 679,width=2.5cm]{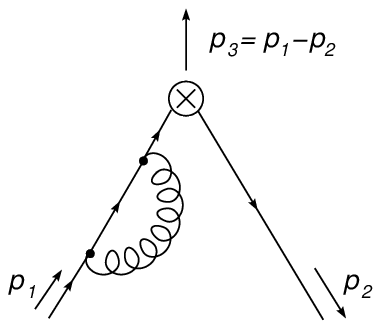}
\vspace{-1cm}
\begin{center}
{\scriptsize{(b)}}
\end{center}
\end{minipage}
\begin{minipage}{2.5cm}
\includegraphics[bb=118 588 226 679,width=2.5cm]{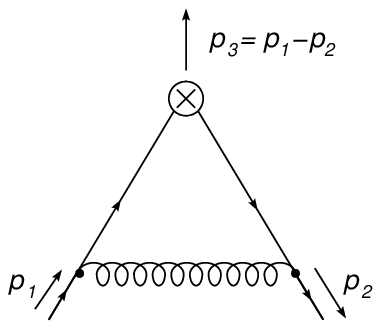}
\vspace{-1cm}
\begin{center}
{\scriptsize{(c)}}
\end{center}
\end{minipage}
\begin{minipage}{2.5cm}
\includegraphics[bb=118 588 226 679,width=2.5cm]{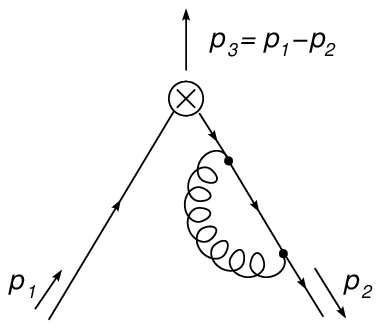}
\vspace{-1cm}
\begin{center}
{\scriptsize{(d)}}
\end{center}
\end{minipage}
\end{center}
\caption{\label{fig:vertex} Diagrams which contribute to the computation
  of the Green's function $G_{\hat{O}}$ at leading order (first line)
  and next-to-leading order(second line) in perturbative QCD. The circle
  with the cross denotes the inserted operator $\hat{O}$. Solid lines
  indicate fermions and spiral lines gluons.}
\end{figure}

The vector and axial-vector Ward-Takahashi identities relate the
self-energies and amputated Green's functions
\begin{equation}
\Lambda_{\hat{O}}=S^{-1}(p_2)\*G_{\hat{O}}\*S^{-1}(p_1)
\end{equation}
of quark bilinear operators $\hat{O}=\bar{u}\Gamma_{\hat{O}} d$ with the
insertion of, in general, the vector($\Gamma_V=\gamma^{\mu}$),
axial-vector($\Gamma_A=\gamma^{\mu}\gamma_5$),
scalar($\Gamma_S=\unitop$) or pseudo-scalar($\Gamma_P=i\*\gamma_5$)
operators.  Diagrams of these Green's functions are shown in
Fig.~\ref{fig:vertex}.
%%%

%

For each of the operators a renormalization constant $Z_{\hat{O}}$ needs
to be introduced with $\hat{O}_R=Z_{\hat{O}}\hat{O}_B$. In the $\MSbar$,
$\ARI$ and $\RIp$ schemes the renormalization constants are related
among each other with $Z_V=1=Z_A$, $Z_P=1/Z_m$ and $Z_S=Z_P$; more
details can be found in
e.g.~Refs.~\cite{Chetyrkin:1996ia,Blum:2001sr}. Some of these properties
hold non-perturbatively while others are proven only in continuum
perturbation theory. Through the Ward-Takahashi identities the
renormalization constants in the $\ARI$ scheme can also be determined by
the conditions on the amputated Green's functions as given in the last
row of Table~\ref{tab:Cond}.  In this case the subtraction point is
exceptional, $p_1^2=p_2^2=-\mu^2$, $p_3=0$, without any momentum
transfer leaving the operator. In Ref.~\cite{Sturm:2009kb} a symmetric
subtraction point has been studied and its applicability demonstrated
through the $\RISMOM$ and $\RISMOMB$ schemes. The conditions for these
new schemes are also summarized in Table~\ref{tab:Cond} and the proofs
for the above relations among the renormalization constants has been
performed in Ref.~\cite{Sturm:2009kb}. A non-perturbative test of the
$\RISMOM$ scheme for the quark mass renormalization can be found in
Ref.~\cite{Aoki:lat2008}.
%%%

%
\begin{table*}[!ht]
\begin{tabular}{|l|l|}
\hline
$\RISMOM$&
$
\lim\limits_{m_R\to0}\left.{1\over12\*q^2}\Tr\left[
q_{\mu}\Lambda^{\mu}_{V,R}(p_1,p_2)\*\Fslash{q}
\right]\right|_{sym}\!\!\!\!\!=1,\quad
\lim\limits_{m_R\to0}\left.{1\over12\*q^2}\Tr\left[
q_{\mu}\Lambda^{\mu}_{A,R}(p_1,p_2)\*\gamma_5\*\Fslash{q}
\right]\right|_{sym}\!\!\!\!\!\!=1
$,\\&
$
\lim\limits_{m_R\to0}\left.{1\over12}\Tr\left[\Lambda_{S,R}(p_1,p_2)\unitop\right]\right|_{sym}\!\!\!\!\!\!=1,\quad
\lim\limits_{m_R\to0}\left.{1\over12\*i}\Tr\left[\Lambda_{P,R}(p_1,p_2)\gamma_5\right]\right|_{sym}\!\!\!\!\!\!=1
$,\\&
%%%%
$
\lim\limits_{m_R\to0}\left.{1\over12\*p^2}\Tr[S_R^{-1}(p)\*\Fslash{p}]\right|_{p^2=-\mu^2}=-1
$,\\&
$
\lim\limits_{m_R\to0}{1\over12\*m_R}\*\left\{
\left.\Tr\left[S^{-1}_R(p)\right]\right|_{p^2=-\mu^2}
-{1\over2}\*\left.\Tr\left[q_{\mu}\Lambda^{\mu}_{A,R}(p_1,p_2)\gamma_5\right]\right|_{sym}
\right\}=1
$.\\ \hline%%%%%%%%%%%%%%%%%%%%%%%%%%%%%%%%%%%%%%%%%%%%%%%%%%%%%%%%%%%%
$\RISMOMB$&
$
\lim\limits_{m_R\to0}\left.{1\over48}\Tr\left[
\Lambda^{\mu}_{V,R}(p_1,p_2)\*\gamma_{\mu}
\right]\right|_{sym}\!\!\!\!\!=1,\qquad
\lim\limits_{m_R\to0}\left.{1\over48}\Tr\left[
\Lambda^{\mu}_{A,R}(p_1,p_2)\*\gamma_5\*\gamma_{\mu}
\right]\right|_{sym}\!\!\!\!\!\!=1
$,\\&
$
\lim\limits_{m_R\to0}\left.{1\over12}\Tr\left[\Lambda_{S,R}(p_1,p_2)\unitop\right]\right|_{sym}\!\!\!\!\!\!=1,\quad
\lim\limits_{m_R\to0}\left.{1\over12\*i}\Tr\left[\Lambda_{P,R}(p_1,p_2)\gamma_5\right]\right|_{sym}\!\!\!\!\!\!=1
$, \\ &
%%%%%
$
\lim_{m_R\to0}\left.{1\over48}\left\{
\Tr\left[\gamma^{\mu}{\partial S_R^{-1}(p)\over\partial p^{\mu}}\right]\right|_{p^2=-\mu^2}
+\Tr\left.\left[
q_{\mu}\gamma^{\alpha}{\partial\over\partial q^{\alpha}}\Lambda^{\mu}_{V,R}
                                 \right]\right|_{sym}
\right\}
=-1
$,\\&
$
\lim_{m_R\to0}{1\over12\*m_R}\*
\left\{
\left.\Tr\left[S^{-1}_R(p)\right]\right|_{p^2=-\mu^2}
-{1\over2}\*\left.\Tr\left[q_{\mu}\Lambda^{\mu}_{A,R}(p_1,p_2)\gamma_5\right]\right|_{sym}
\right\}
=\phantom{+}1
$.\\\hline\hline%%%%%%%%%%%%%%%%%%%%%%%%%%%%%%%%%%%%%%%%%%%%%%%%%%%%%%%%%%%%
$\ARI$&
$
\lim\limits_{m_R\to0}\left.{1\over48}\Tr\left[
\Lambda^{\mu}_{V,R}(p_1,p_2)\*\gamma_{\mu}
\right]\right|_{asym}\!\!\!\!\!=1,\quad
\lim\limits_{m_R\to0}\left.{1\over48}\Tr\left[
\Lambda^{\mu}_{A,R}(p_1,p_2)\*\gamma_5\*\gamma_{\mu}
\right]\right|_{asym}\!\!\!\!\!\!=1
$,\\&
$
\lim\limits_{m_R\to0}\left.{1\over12}\Tr\left[\Lambda_{S,R}(p_1,p_2)\unitop\right]\right|_{asym}\!\!\!\!\!\!=1,\quad
\lim\limits_{m_R\to0}\left.{1\over12\*i}\Tr\left[\Lambda_{P,R}(p_1,p_2)\gamma_5\right]\right|_{asym}\!\!\!\!\!\!=1
$,\\ &
%%%
$
\lim\limits_{m_R\to0}\left.{1\over48}\Tr\left[\gamma^{\mu}{\partial
    S_R^{-1}(p)\over\partial p^{\mu}}\right]\right|_{p^2=-\mu^2}=-1,
\quad
\lim\limits_{m_R\to0}\left.{1\over12\*m_R}\Tr[S_R^{-1}(p)]\right|_{p^2=-\mu^2}=1
$.\\\hline%%%%%%%%%%%%%%%%%%%%%%%%%%%%%%%%%%%%%%%%%%%%%%%%%%%%%%%%%%%%
\end{tabular}
\caption{\label{tab:Cond} This table summarizes the renormalization
  conditions for the $\RISMOM$ and $\RISMOMB$ schemes as defined in
  Ref.~\cite{Sturm:2009kb} for the amputated Green's functions of the
  vector, axial-vector, scalar and pseudo-scalar operators as well as
  for the self-energies. The last row contains the conditions for the
  $\ARI$ scheme~\cite{Martinelli:1994ty}. The subscript $(a)sym$ stands
  for the restriction to the (a)symmetric momentum configuration and
  $q=p_1-p_2$ is the momentum transfer leaving the operator.}
\end{table*}
Also the tensor operator, with
$\Gamma_T=\sigma^{\mu\nu}={i\over2}[\gamma^{\mu},\gamma^{\nu}]$, has
been studied in
Refs.~\cite{Broadhurst:1994se,Gracey:2000am,Gracey:2003yr} and in
Ref.~\cite{Sturm:2009kb} for a symmetric subtraction point.
\section{Results\label{sec:results}}
\subsection{Matching factors}
The matching factors which convert a quark mass from the $\RISMOM$ or
$\RISMOMB$ scheme to the $\MSbar$ scheme
\begin{equation}
m^{\msbar}_R=C^{\mbox{\scriptsize RI/SMOM}_{(\gamma_\mu)}}_m\,m^{\mbox{\scriptsize RI/SMOM}_{(\gamma_\mu)}}_R 
\end{equation}
have been computed to one-loop order in perturbative QCD in
Ref.~\cite{Sturm:2009kb}. The numerically evaluated result allows for a
direct comparison of the size of the perturbative coefficients in the
different schemes. In the $\RISMOM$ scheme the result reads
\begin{eqnarray}
C^{\rismom}_m&=&1-{\alpha_s\over4\*\pi}[ 
       0.645518856...\nonumber\\&&\qquad\,
- \xi\*0.229271492...]+\mathcal{O}(\alpha_s^2).
\end{eqnarray}
The corresponding conversion factor for the fermion fields
$C^{\rismom}_q=Z^{\msbar}_q/Z^{\rismom}_q$ in the case of the
$\RISMOM$ scheme is given by
\begin{equation}
C^{\rismom}_q=1-{\alpha_s\over4\*\pi}\*\xi\*1.333333333...+\mathcal{O}(\alpha_s^2).
\end{equation}
In the $\RISMOMB$ scheme the results for the matching factors read
\begin{eqnarray}
C^{\rismomb}_m\!\!&\!=\!&\!1-{\als\over4\*\pi}[
       1.978852189...\nonumber\\&&\qquad
+ \xi\*0.666666667...]+\mathcal{O}\left(\alpha^2_s\right)
\end{eqnarray}
and
\begin{eqnarray}
C^{\rismomb}_q\!\!&\!=\!&\!1+{\alpha_s\over4\*\pi}\*[
       1.333333333...\nonumber\\&&\qquad
- \xi\*0.437395174...]+\mathcal{O}\left(\alpha^2_s\right)\!\!,\;
\end{eqnarray}
where $\xi$ is the gauge parameter and $\alpha_s$ the strong coupling
constant.
%%%

In order to determine the perturbative mass conversion factors $C_m$ for
a symmetric subtraction point in the two mentioned schemes, one can
consider the computation of the amputated Green's function of the
(pseudo-)scalar operator or the (axial-)vector operator with suitable
projectors. In the case of the pseudo-scalar and axial-vector operator a
naive anti-commuting definition of $\gamma_5$ is used for the treatment
of $\gamma_5$ in dimensional
regularization~\cite{tHooft:1972fi,Breitenlohner:1977hr} which is a
self-consistent prescription for the flavor non-singlet
contributions~\cite{Trueman:1979en,Larin:1993tq}.
%%%

Comparing these matching factors with the conversion factor in the
$\ARI$ scheme
\begin{eqnarray}
C^{\ari}_m&=&1-{\alpha_s\over4\*\pi}\*[
       5.333333333...\nonumber\\&&\qquad
+ \xi\*2.000000000...]+\mathcal{O}\left(\alpha^2_s\right)
\end{eqnarray}
and the $\RIp$ scheme\footnote{In the Landau gauge at one-loop order the
conversion factors $C_m$ are the same for the $\ARI$ and $\RIp$
schemes.}
\begin{eqnarray}
C^{\rip}_m\!&\!=\!&\!1-{\alpha_s\over4\*\pi}\*[
   5.333333333...\nonumber\\&&\qquad
+ \xi\*1.333333333...]+\mathcal{O}\left(\alpha^2_s\right)\!\!,\;
\end{eqnarray}
one finds in particular in the Landau gauge ($\xi=0$), which is usually
adopted in the lattice simulations, a significantly smaller one-loop
coefficient which suggests a better behaved perturbative series. If this
behavior persists at higher orders it allows for a substantial reduction
of the systematic uncertainty in the light quark mass determinations.
%%%

\begin{figure}[!ht]
\begin{minipage}{8.5cm}
\begin{center}
\includegraphics[bb=0 0 567 384,width=8.5cm]{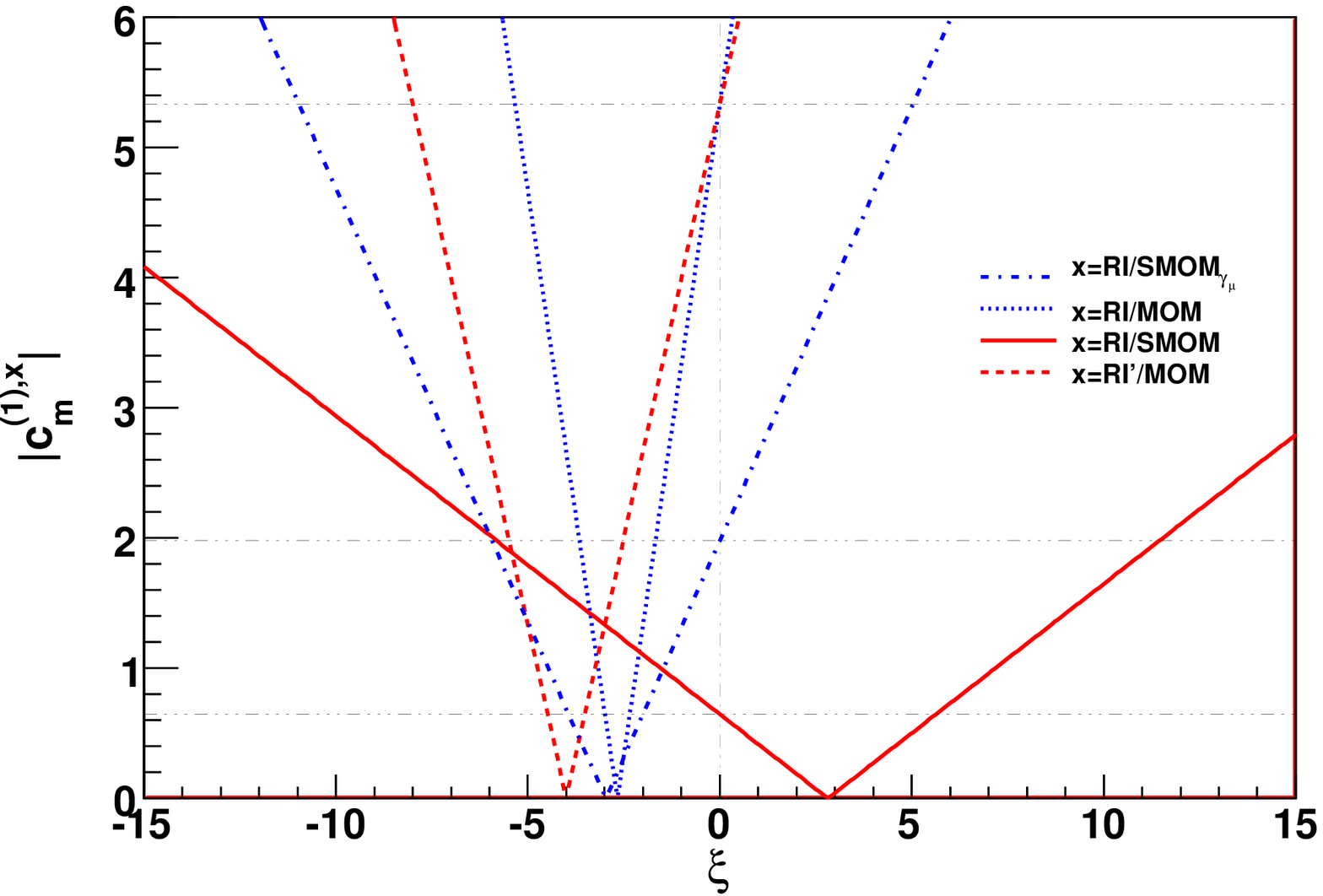}
\end{center}
\end{minipage}
\caption{\label{fig:cm} This figure shows the one-loop coefficients
  $c^{(1),x}_m$ of the matching factors $C^{x}_m$ as a function of the
  gauge parameter $\xi$ for two exceptional (dashed line(red), $x=\RIp$;
  dotted line(blue), $x=\ARI$) and two non-exceptional (solid line(red),
  $x=\RISMOM$; dashed-dotted line(blue), $x=\RISMOMB$) momentum
  configurations. The horizontal lines indicate the size of the one-loop
  coefficient in the Landau gauge($\xi=0$).}
\end{figure}

The matching factors $C^{x}_{m}$ in the $x=\ARI$, $x=\RIp$, $x=\RISMOM$
and $x=\RISMOMB$ schemes are gauge dependent. While performing the
conversion of the quark mass to the $\MSbar$ scheme this gauge
dependence is compensated by the gauge dependence of the renormalization
constant determined in the lattice simulation. The one-loop coefficient
$c^{(1),x}_m(\xi)$ of the matching factor
$C^{x}_m=1+{\als\over4\*\pi}\*c^{(1),x}_m(\xi)$ as a function of $\xi$
is shown in Fig.~\ref{fig:cm} for the exceptional and non-exceptional
momentum configurations.
For almost all gauges the one-loop coefficient of the symmetric
subtraction point is smaller than in the asymmetric case except for a
small window with $\xi\in(-5.41532...,-3)$ comparing the $\RIp$ and
$\RISMOM$ schemes and $\xi\in(-2.74207...,-2.51586...)$ comparing the
$\ARI$ and $\RISMOMB$ schemes.
\subsection{Two-loop anomalous dimensions}
The mass anomalous dimension
\begin{eqnarray}
\gamma_m&=&{d\log m(\mu)\over d\log\left(\mu^2\right)}=
-\gamma^{(0)}_m{\alpha_s\over\pi}
-\gamma^{(1)}_m\*\left({\alpha_s\over\pi}\right)^2\nonumber\\
&-&\gamma^{(2)}_m\*\left({\alpha_s\over\pi}\right)^3
+\mathcal{O}(\alpha^4_s)
\end{eqnarray}
can be used to run the quark mass to different energy scales. In the
$\MSbar$ scheme the lowest expansion coefficients read
\begin{eqnarray}
\gamma_m^{(0),\msbar}\!&\!\!=\!\!&\!{3\over4}\*\CF,\\
\gamma_m^{(1),\msbar}\!&\!\!=\!\!&
\!{1\over16}\!\left({3\over2}\*\CF^2+{97\over6}\*\CF\*\CA-{10\over3}\*\CF\*\TF\*\nf\right)\!\!,\,\;
\end{eqnarray}
where $C_F=4/3$($C_A=3$) denotes the Casimir operator in the
fundamental(adjoint) representation of SU(3). The symbol $n_f$
represents the number of active fermions and $T_F=1/2$ is the
normalization of the trace of the SU(3) generators in the fundamental
representation.
%%%

The mass anomalous dimensions of the $\RISMOM$ and $\RISMOMB$ schemes up
to two-loop order are given by~\cite{Sturm:2009kb}
\begin{eqnarray}
\gamma^{(0),\rismom}_m&=&\gamma^{(0),\msbar}_m=\gamma^{(0),\rismomb}_m,\\
\gamma^{(1),\rismom}_m&=&\gamma^{(1),\msbar}_m\nonumber\\
                      &+&\beta^{(0)}\*\CF\* 0.121034785...,\\
\gamma^{(1),\rismomb}_m&=&\gamma^{(1),\msbar}_m\nonumber\\
                       &+&\beta^{(0)}\*\CF\*0.371034785...,
\end{eqnarray}
with the QCD $\beta$-function defined by
\begin{equation}
\beta={d\alpha_s(\mu)/\pi\over d\log(\mu^2)}=
-\beta^{(0)}\*\left({\alpha_s\over\pi}\right)^2
+\mathcal{O}(\alpha^3_s)
\end{equation}
and the lowest order coefficient
\begin{equation}
\beta^{(0)}={1\over4}\*\left({11\over3}\*\CA-{4\over3}\*\TF\*\nf\right).
\end{equation}
%%%

Similarly one can define the anomalous dimension
\begin{equation}
\gamma_q=2{d\log\Psi_R\over d\log(\mu^2)}.
\end{equation}
As shown in Ref.~\cite{Sturm:2009kb} it is equal in the $\RISMOM$ and
$\RIp$ schemes and known up to order $\alpha_s^3$ in
Refs.~\cite{Chetyrkin:1999pq,Gracey:2003yr}. In the $\RISMOMB$ scheme it
is given in the Landau gauge by
\begin{eqnarray}
\gamma^{}_q&=&\left({\alpha_s\over\pi}\right)^2\*
\left({3\over32}\*\CF^2 - {31\over192}\*\CF\*\CA\right.\nonumber\\
&&\qquad\qquad\left.+ {1\over24}\*\CF\*\TF\*n_f\right)+\mathcal{O}(\alpha_s^3)
\end{eqnarray}
and has been determined in Ref.~\cite{Sturm:2009kb}.
\section{Conclusions\label{sec:conclusion}}
We have discussed the renormalization of quark masses in a
regularization-invariant momentum subtraction scheme with a symmetric
subtraction point. The scheme is useful in the context of lattice
simulations of the light up-, down- and strange-quark masses as an
intermediate step before the conversion to the $\MSbar$ scheme. The
renormalization constants can be determined by the computation of
amputated Green's functions with the insertion of the scalar,
pseudo-scalar, vector and axial-vector operators. This allows one to
derive in continuum perturbation theory the two-loop anomalous
dimensions and one-loop matching factors which convert the light quark
masses renormalized in the $\RISMOM$ scheme into the $\MSbar$
scheme. The new $\RISMOM$ scheme decreases chiral symmetry breaking as
well as other unwanted infrared effects in the lattice calculation and
the one-loop coefficient of the new matching factor is smaller compared
to the one of the $\ARI$($\RIp$) scheme.  A better behaved perturbative
series can improve the accuracy of light quark mass determinations, if
this behavior is confirmed at higher orders.
%%%

%%%%%%%%%%%%%%%%%%%%%%%%%%%%%%%%%%
\begin{acknowledgments}
I would like to thank Y. Aoki, N.H. Christ, T. Izubuchi, C.T Sachrajda
and A. Soni for the fruitful collaboration on this subject. This work was
supported by U.S. DOE under contract No.DE-AC02-98CH10886. 
\end{acknowledgments}

%%%%%%%%%%%%%%%%%%%%%%%%%%%%%%%%%%
\bigskip %
\end{document}